\documentclass{ptephy_v1}

\preprintnumber{XXXX-XXXX} 
\usepackage{hyperref}

\usepackage{graphics} 

\begin{document}

\title{On the mechanism of black hole energy reduction in the Blandford-Znajek process}


\author[1,2]{Kenji Toma}
\author[3]{Fumio Takahara}
\author[4,5]{Masanori Nakamura}
\affil[1]{Frontier Research Institute for Interdisciplinary Sciences, Tohoku University, Sendai 980-8578, Japan\email{toma@astr.tohoku.ac.jp}}
\affil[2]{Astronomical Institute, Tohoku University, Sendai 980-8578, Japan}
\affil[3]{Department of Earth and Space Science, Graduate School of Science, Osaka University, Toyonaka 560-0043, Japan}
\affil[4]{National Institute of Technology, Hachinohe College, 16-1 Uwanotai, Tamonoki, Hachinohe, Aomori 039-1192, Japan}
\affil[5]{Institute of Astronomy and Astrophysics, Academia Sinica, P.O. Box 23-141, Taipei 10617, Taiwan, R.O.C.}




\begin{abstract}
  The Blandford-Znajek (BZ) process is steady electromagnetic energy release from rotating black holes (BHs) along magnetic field lines threading them and widely believed to drive relativistic jets. This process is successfully demonstrated in general relativistic magnetohydrodynamic (MHD) simulations with the coordinate system regular on the event horizon, in which the outward Poynting flux on the horizon is considered to reduce BH energy. Meanwhile, alternative pictures for the BH energy reduction that invoke infall of negative energy objects were also discussed, although all of the proposed definitions of the negative energy and/or its infall velocity were ambiguous.
  We revisit the mechanism of BH energy reduction in the BZ process under the ideal MHD condition by utilizing the coordinate system singular on the horizon, in which the falling membrane of past accreted matter should exist above the horizon. We find that the Poynting flux is produced at the boundary between the falling membrane and the magnetically-dominated inflow, and the front of the inflow creates the negative electromagnetic energy, which reduces the rotational energy of spacetime. We also clarify that the poloidal electric current does not form a closed circuit within the magnetically-dominated flow. Previous interpretations of the BZ process and possibilities of violation of ideal MHD condition and BH charging are also discussed.
\end{abstract}

\subjectindex{xxxx, xxx}

\maketitle

\section{Introduction}
\label{sec:intro}

Relativistic jets are collimated outflows with velocities close to the light speed, which are observed in active galactic nuclei, gamma-ray bursts, galactic microquasars, and so on. Their main driving mechanism is widely believed to be the Blandford-Znajek (BZ) process \cite{blandford77}, i.e., the steady electromagnetic extraction of black hole (BH) rotational energy. General relativistic magnetohydrodynamic (GRMHD) simulations have been developed since 2000s \cite{komissarov01,koide02,komissarov04b}, and the simulations of radiatively inefficient accretion flows (RIAFs) have shown that BH magnetospheres can form, where outward Poynting flux is maintained along the ordered magnetic field lines threading the event horizon via the BZ process and fluid is electromagnetically accelerated to the relativistic speed \cite[e.g.][]{mckinney04,barkov08,tchekhovskoy11,takahashi16,nakamura18,porth19,liska20,narayan22,ripperda22}. This picture is supported by very long baseline interferometric (VLBI) observations of the jet of M87, which is one of the closest radio galaxies \cite[e.g.][]{nakamura18,hada11,eht19a,eht19b,eht21,eht24}.

The BZ process was theoretically discovered by analogy with the unipolar induction process of rotating neutron stars with dipole magnetic field \cite{blandford77,goldreich69,wald74}.
The neutron star is matter-dominated, where the poloidal electric current is driven towards the higher electric potential by the rotation of matter, producing the Poynting flux and reducing the rotational energy of matter \cite{beskin10}\cite[see also][]{shibata21}. In contrast, there is no matter-dominated region in the BZ process. The Poynting flux appears to directly reduce the BH rotational energy without effective exchange between matter and electromagnetic energies. This is clearly demonstrated in the GRMHD simulations with coordinate system regular on the event horizon, Kerr-Schild (KS) coordinates.\footnote{The GRMHD simulations fix the spacetime metric. The decrease of BH spin is actually shown by solving Einstein equation with an analytical perturbative method \cite{kimura21} and with numerical relativity simulations \cite{shibata24}.}

The outward Poynting flux on the horizon
in the GRMHD simulations
is possible because this flux vector is space-like in the ergosphere, and
no physical observer measures this flux \cite{komissarov09,lasota14}.
The outward Poynting flux is measurable by an observer at the far zone. Some theorists
tried to find an alternative picture that invokes infall of negative energy objects in a similar manner to the mechanical Penrose process \cite{penrose69}, but all of the proposed definitions of the negative energy and/or its infall velocity were ambiguous \cite{takahashi90,komissarov09,koide14} (see Section~\ref{sec:literature}).

In this paper, we provide a clear picture 
for the mechanism of BH energy reduction with negative electromagnetic energy in the BZ process 
by using the cooridate system singular on the event horizon, Boyer-Lindquist (BL) coordinates. For the basis of this picture, we show analytical GRMHD solutions along a magnetic field line near the horizon.
A novel point in our picture
is that we treat a falling membrane of past accreted matter that should exist above the horizon. Then we find that the Poynting flux is produced outside the event horizon, associated with production of negative electromagnetic energy. We also clarify that the poloidal electric current that maintains the Poynting flux cannot form a closed circuit within the magnetosphere. Such a clarification may be useful for gaining insights into detailed plasma physics close to the horizon.

In Section~\ref{sec:review}, we briefly review the description of particles and electromagnetic fields in the Kerr spacetime and the formulation of steady, axisymmetric Kerr BH magnetospheres.
In Section~\ref{sec:mhd}, we show analytical solutions of steady, axisymmetric GRMHD inflow along a magnetic field line. Based on these knowledges, we provide our picture for the BH energy reduction and the electric current closure in Section~\ref{sec:picture}. We discuss the previous interpretations of the BZ process and the driving force of poloidal current in Section~\ref{sec:discussion}.

\section{Black hole magnetospheres}
\label{sec:review}

\subsection{The Kerr spacetime and particles near the horizon}

The steady, axisymmetric spacetime of a rotating BH is described by the Kerr metric, written in the BL coordinates $(t, \varphi, r, \theta)$ as
\begin{equation}
ds^2 = g_{\mu\nu}dx^\mu dx^\nu = - \alpha^2 dt^2 + \gamma_{\varphi\varphi}(d\varphi - \Omega dt)^2 + \gamma_{rr}dr^2 + \gamma_{\theta\theta}d\theta^2,
\end{equation}
where
\begin{equation}
\alpha = \sqrt{\frac{\varrho^2 \Delta}{\Sigma}}, ~~~ \Omega = \frac{2r}{\Sigma}a, ~~~ \gamma_{\varphi\varphi} = \frac{\Sigma}{\varrho^2}\sin^2\theta, ~~~ \gamma_{rr} = \frac{\varrho^2}{\Delta}, ~~~ \gamma_{\theta\theta} = \varrho^2,
\end{equation}
with $\varrho^2 = r^2 + a^2\cos^2\theta$, $\Delta = r^2 + a^2 -2r$, and $\Sigma = (r^2+a^2)^2 -a^2\Delta\sin^2\theta$. We treat fixed BH mass $M$ and angular momentum $J$ and consider the matter and electromagnetic fields as test fields. We have adopted the units of $c=1$ and $GM=1$, for which the gravitational radius $r_g = GM/c^2 = 1$, and used the dimensionless spin parameter $a = J/(M r_g c)$.
The BL coordinates have a well-known coordinate singularity at the event horizon, $r = 1+\sqrt{1-a^2} \equiv r_{\rm H}$, where $\gamma_{rr} = \infty$, $\Delta = 0$ and $\alpha = 0$.

The ergosphere is the region of $r < 1+\sqrt{1-a^2\cos^2\theta} \equiv r_{\rm es}$, where the Killing vector $\xi^\mu = (1, 0, 0, 0)$ is space-like, $\xi^2 = g_{tt} = -\alpha^2 + \gamma_{\varphi\varphi}\Omega^2 > 0$. Thus, the world line of constant $(\varphi, r, \theta)$ is space-like in the ergosphere, and time-like particle motions are restricted to $d\varphi > 0$ (`frame-dragging effect'). The particle energy per mass $-u_\mu \xi^\mu = -u_t$ can be negative in the ergosphere. The reduction of BH rotational energy via infalling of a particle with $-u_t < 0$ is called mechanical Penrose process \cite{penrose69}.

The fiducial observers (FIDOs) in the BL coordinates, described by the world line perpendicular to the hypersurface of constant $t$, $d\varphi = \Omega dt, dr = d\theta = 0$, are time-like everywhere (as long as outside the event horizon) \cite{thorne86}. The FIDOs have proper time $d\tau = \alpha dt$ and coordinate four-velocity
\begin{equation}
  n^\mu = \left(\frac{1}{\alpha}, \frac{\Omega}{\alpha}, 0, 0 \right), ~~~
  n_\mu = g_{\mu\nu} n^\nu = (-\alpha, 0, 0, 0).
\end{equation}
The FIDOs rotate in the same direction as the BH with the coordinate angular velocity $d\varphi/dt = \Omega$, but have zero angular momentum, $n_\mu \chi^\mu = n_\varphi = 0$, where $\chi^\mu = (0, 1, 0, 0)$ is also the Killing vector. The angular velocity at the event horizon $\Omega(r_{\rm H}) = a/(2r_{\rm H}) \equiv \Omega_{\rm H}$ is interpreted as the BH angular velocity.

The FIDOs can have a local orthonormal basis,
\begin{eqnarray}
\omega^{(t)} = \alpha dt, ~~~ \omega^{(\varphi)} = \sqrt{\gamma_{\varphi\varphi}}(d\varphi - \Omega dt), ~~~ \omega^{(r)} = \sqrt{\gamma_{rr}}dr, ~~~ \omega^{(\theta)} = \sqrt{\gamma_{\theta\theta}}d\theta,
\end{eqnarray}
and its dual basis $e_{(t)} = (1/\alpha)(\partial_t +\Omega\partial_\varphi)$, $e_{(\varphi)} = (1/\sqrt{\gamma_{\varphi\varphi}})\partial_\varphi$, $e_{(r)}=(1/\sqrt{\gamma_{rr}})\partial_r$, $e_{(\theta)}=(1/\sqrt{\gamma_{\theta\theta}})\partial_\theta$ \cite[e.g.][]{thorne86}. The four-velocity components of a particle with respect of the orthonormal basis are calculated as $u^{\hat{t}} = u^\mu \omega_\mu^{(t)} = \alpha u^t$, $u^{\hat{\varphi}} = \sqrt{\gamma_{\varphi\varphi}}(u^\varphi - \Omega u^t)$, so that the FIDO measures the toroidal velocity as $v^{\hat{\varphi}} = u^{\hat{\varphi}}/u^{\hat{t}} = (\sqrt{\gamma_{\varphi\varphi}}/\alpha)(v^\varphi - \Omega)$.
Since $|v^{\hat{\varphi}}| < 1$, a particle's coordinate toroidal velocity
\begin{equation}
  v^\varphi \to \Omega_{\rm H},
  \label{eq:ff_vphi}
\end{equation}
as it approaches the horizon $(\alpha \to 0)$.

For a freely falling particle close to the horizon, we have Eq.~(\ref{eq:ff_vphi}) as well as
\begin{equation}\label{eq:v_ff}
  v^r \to -\frac{\alpha}{\sqrt{\gamma_{rr}}},
\end{equation}
by integration of geodesic equation \cite{carter68,punsly08}. This corresponds to $v^{\hat{r}} \to -1$ in the FIDO frame, which results from the divergently strong gravitational force (proportional to $\partial_r \ln \alpha$ \cite{thorne86,levinson18}). The energy per mass of the particle diverges as $-u_{\hat{t}} = -u_\mu e^\mu_{(t)} = -(u_t + \Omega u_\varphi)/\alpha \propto \alpha^{-1}$, where we note that $u_t = u_\mu \xi^\mu$ and $u_\varphi = u_\mu \chi^\mu$ are constant along the geodesic.

\subsection{The 3+1 electrodynamics}

The covariant Maxwell equations $\nabla_\nu{}^*F^{\mu\nu}=0$ and $\nabla_\nu F^{\mu\nu} = 4\pi I^\nu$ can be reduced by using four spatial vectors $D^\mu = F^{\mu\nu}n_\nu$, $B^\mu = -{}^*F^{\mu\nu}n_\nu$, $E^\mu = F^{\mu\nu}\xi_\nu$, and $H^\mu = -{}^*F^{\mu\nu}\xi_\nu$ to \cite{komissarov04a,landau75}
\begin{equation}\label{eq:maxwell1}
  \nabla \cdot \mathbf{B} = 0, ~~~ \frac{\partial}{\partial t} \mathbf{B} + \nabla \times \mathbf{E} = 0,
\end{equation}
\begin{equation}\label{eq:maxwell2}
  \nabla \cdot \mathbf{D} = 4\pi \rho_e, ~~~ -\frac{\partial}{\partial t} \mathbf{D} + \nabla \times \mathbf{H} = 4\pi \mathbf{J},
\end{equation}
where $\nabla \cdot \mathbf{C}$ and $\nabla \times \mathbf{C}$ denote $(1/\sqrt{\gamma})\partial_i (\sqrt{\gamma} C^i)$ and $(1/\sqrt{\gamma}) \epsilon^{ijk} \partial_j C_k$, respectively, and $\epsilon^{ijk}$ is the Levi-Civita pseudo-tensor. $D^\mu$ and $B^\mu$ are the electric and magnetic fields as measured by the FIDOs, while $E^\mu$ and $H^\mu$ are those in the coordinate basis. These sets of electromagnetic fields have constitutive relations,
\begin{equation} \label{eq:consti}
  \mathbf{E} = \alpha \mathbf{D} - \Omega\mathbf{m} \times \mathbf{B},~~~
  \mathbf{H} = \alpha \mathbf{B} + \Omega\mathbf{m} \times \mathbf{D},
\end{equation}
where $\mathbf{m} = \partial_\varphi$ and $\mathbf{C} \times \mathbf{F}$ denotes $(1/\sqrt{\gamma})\epsilon^{ijk} C_i F_k$.
The charge density and the electric current in Eq.~(\ref{eq:maxwell2}) are defined by $\rho_e = -I^\mu n_\mu$ and $J^\mu = (\xi^\mu I^\nu - \xi^\nu I^\mu)n_\nu$.

The covariant energy-momentum equation of electromagnetic field $\nabla_\nu T^\nu_{~\mu} = -F_{\mu\nu} I^\nu$ gives us the energy and angular momentum equations,
\begin{equation}\label{eq:e_cons}
  \frac{\partial}{\partial t} e + \nabla \cdot \mathbf{S} = -\mathbf{E}\cdot\mathbf{J},
\end{equation}
\begin{equation}\label{eq:am_cons}
  \frac{\partial}{\partial t} \ell + \nabla\cdot \mathbf{L} = -(\rho \mathbf{E} + \mathbf{J}\times\mathbf{B})\cdot \mathbf{m}, 
\end{equation}
where $\mathbf{C} \cdot \mathbf{F} = C^iF_i$. The energy density $e ~(= -\alpha T^t_{~t})$ and the Poynting flux $S^i ~(= -\alpha T^i_{~t})$ are written by
\begin{equation}\label{eq:EMenergy}
  e = \frac{1}{8\pi} (\mathbf{E}\cdot\mathbf{D} + \mathbf{B}\cdot\mathbf{H}),
\end{equation}
\begin{equation}
  \mathbf{S} = \frac{1}{4\pi}\mathbf{E}\times\mathbf{H}.
\end{equation}
The angular momentum density and the angular momentum flux are written by $\ell ~(= \alpha T^t_{~\varphi}) = (1/4\pi)(\mathbf{D}\times\mathbf{B})\cdot \mathbf{m}$ and $L^i ~(=\alpha T^i_{~\varphi}) = -(1/4\pi)(\mathbf{E}\cdot\mathbf{m})\mathbf{D} - (1/4\pi)(\mathbf{H}\cdot\mathbf{m})\mathbf{B} + (1/8\pi)(\mathbf{E}\cdot\mathbf{D}+\mathbf{B}\cdot\mathbf{H})\mathbf{m}$, respectively.

The electromagnetic fields measured by the FIDOs generally obey the regularity condition at the event horizon (`Znajek condition'),
\begin{equation}\label{eq:znajek}
D^{\hat{\theta}} = -B_{\hat{\varphi}}.
\end{equation}
This condition is derived from the condition that the fields in the coordinates regular on the horizon do not diverge on the horizon \cite{znajek77,dermer09}.

\subsection{Steady, axisymmetric magnetospheres and the BZ solution}
\label{sec:ff}

In this paper we basically consider steady, axisymmetric magnetospheres of Kerr BHs, in which the electromagnetic energy density dominates the plasma particle energy density, and the event horizon is threaded by the ordered magnetic field produced by the electric current of plasma.
We consider the BZ process as the steady production of outward Poynting flux along such ordered magnetic field lines.\footnote{It can be generically proved that the BZ process operates, i.e., $\Omega_{\rm F} > 0$ and $H_\varphi \neq 0$, when the magnetic field lines are threading the ergosphere and $\mathbf{D}\cdot\mathbf{B} = 0$ \citep{komissarov04a,toma14}. Toma and Takahara \citep{toma14} argued that the field lines need to cross the outer light surface, but only crossing of inner light surface is sufficient.} 

We assume that the particle number density is high enough to screen the electric field along the magnetic field lines,
\begin{equation}\label{eq:ff}
  \mathbf{D} \cdot \mathbf{B} = 0.
\end{equation}
This condition and Eq.~(\ref{eq:consti}) lead to $\mathbf{E}\cdot\mathbf{B}=0$. In the steady state, we have $\nabla \times \mathbf{E}=0$, which means that $\mathbf{E}$ is a potential field, and the axisymmetry leads to $E_\varphi = 0$. Then one can write
\begin{equation}\label{eq:omegaF}
  \mathbf{E} = -\Omega_{\rm F}\mathbf{m} \times \mathbf{B}.
\end{equation}
The so-called `field line angular velocity' $\Omega_{\rm F}$ is found to be constant along each magnetic field line by substituting Eq.~(\ref{eq:omegaF}) into $\nabla \times \mathbf{E} = 0$.
In this case, the poloidal Poynting flux and the poloidal angular momentum flux are written by
\begin{equation}
  \mathbf{S}_{\rm p} = \frac{-H_\varphi \Omega_{\rm F}}{4\pi}\mathbf{B}_{\rm p}, ~~~
  \mathbf{L}_{\rm p} = \frac{-H_\varphi}{4\pi}\mathbf{B}_{\rm p}.
\end{equation}
The angular momentum equation (Eq.~\ref{eq:am_cons}), $\nabla \cdot (-H_\varphi\mathbf{B}_{\rm p}/4\pi) = B^i \partial_i(-H_\varphi/4\pi) = -(\mathbf{J}_{\rm p}\times \mathbf{B}_{\rm p})\cdot\mathbf{m}$, implies that $H_\varphi = {\rm const.}$ along a magnetic field line if the poloidal electric current does not cross it, $\mathbf{J}_{\rm p} \times \mathbf{B}_{\rm p} = 0$.

Based on the above formulation, Blandford and Znajek \cite{blandford77} further assume the force-free condition, $\rho \mathbf{E} + \mathbf{J} \times \mathbf{B} = 0$, to find perturbative solutions of electromagnetic field in the slow rotation limit $a \ll 1$. 
For deriving the solutions, the regularity condition at the event horizon (Eq.~\ref{eq:znajek}) and the condition at the far zone \cite{michel73}, $B_{\hat{\varphi}} = E_{\hat{\theta}}$, are utilized. They are written by $H_\varphi = (\Omega_{\rm F} - \Omega_{\rm H}) \sqrt{\gamma} B^r \sin\theta$ and $H_\varphi = -\Omega_{\rm F} \sqrt{\gamma}B^r \sin\theta$, respectively, in the case of split-monopole magnetic field at the zeroth order ($B^r \propto 1/\sqrt{\gamma}$ and $B^\theta = B^\varphi = 0$).
The
solution $\Omega_{\rm F} \approx \Omega_{\rm H}/2$ can be understood
by matching the two conditions along each field line with the constancy of $H_\varphi, \Omega_{\rm F}, \sqrt{\gamma}B^r,$ and $\theta$ \citep{komissarov04a}.

It was argued that the BZ solutions of electromagnetic field with the force-free condition have a causality problem, since they are determined by the conditions at the event horizon and at infinity \cite{punsly89}. The horizon must not yield any condition that actively affects its exterior, only passively absorbing particles and waves. This causality problem is avoided in the approach with the MHD condition \cite{takahashi90,levinson04,komissarov04a,beskin10} (see Sections~\ref{sec:mhd} and \ref{sec:picture}).

\section{Steady axisymmetric MHD flows}
\label{sec:mhd}



In this section we investigate the detailed structure of steady, axisymmetric, cold flows in the BZ process under the ideal MHD condition, based on the formalism presented in Section~\ref{sec:ff}. This is required for building a clear picture for the mechanism of BH energy reduction (Section~\ref{sec:picture}).
 
The ideal MHD condition, $F_{\mu\nu} u^\nu = 0$, is reduced to $\mathbf{E} = -\mathbf{v} \times \mathbf{B}$,
so that $\mathbf{E}\cdot\mathbf{B}=0$, and $\Omega_{\rm F}$ is defined by Eq.~(\ref{eq:omegaF}) for steady, axisymmetric flows.
Then one has
\begin{equation}\label{eq:MHD_v}
  v^r = \kappa B^r, ~~~ v^\theta = \kappa B^\theta, ~~~
  v^\varphi = \kappa B^\varphi + \Omega_{\rm F},
\end{equation}
where $\kappa$ is a scalar function. In addition to $\Omega_{\rm F}$, one has three quantities conserved along each magnetic field line from the conservation equations of mass, energy, and angular momentum \cite{bekenstein78},
\begin{equation}\label{eq:MHD_eta}
  \eta = \alpha \rho u^t \kappa,
\end{equation}
\begin{equation}\label{eq:MHD_E}
  \mathcal{E} = -u_t -\frac{\Omega_{\rm F}H_\varphi}{4\pi\eta},
\end{equation}
\begin{equation}\label{eq:MHD_L}
  \mathcal{L} = u_\varphi -\frac{H_\varphi}{4\pi\eta},
\end{equation}
where $\rho$ is the fluid-frame mass density. Given the poloidal field profile, $B^r$ and $B^\theta$, with the conserved quantities, $\Omega_{\rm F}, \mathcal{E},$ and $\mathcal{L}$ for each field line, the above equations can be solved for $u_\mu$ and $H_\varphi$. $\eta$ does not affect the structure of flows (see Appendix~\ref{sec:app}). The flow solution consists of inflow ($\kappa < 0, \eta < 0, \mathcal{E} < 0, \mathcal{L} < 0$) and outflow ($\kappa > 0, \eta > 0, \mathcal{E} > 0, \mathcal{L} > 0$) which are separated at the point where the gravity and Lorentz force are balanced.

$\Omega_{\rm F}, \mathcal{E},$ and $\mathcal{L}$ are determined by the regularity condition at the inner magnetosonic point in the inflow, that at the outer magnetosonic point in the outflow, and the location of the separation point, without the condition at the event horizon \cite{takahashi90,pu20}.
For an inappropriate set of $\Omega_{\rm F}, \mathcal{E},$ and $\mathcal{L}$, one cannot obtain the trans-magnetosonic inflow continuous towards the horizon through the inner magnetosonic point or the trans-magnetosonic outflow continuous towards infinity through the outer magnetosonic point.
To find the poloidal field profile, one must solve the transverse force balance between the magnetic field lines \cite{beskin00,huang20,ogihara21}.

Here we focus on the inflow and present its analytical solutions along a field line in the BL coordinates by using the method of Takahashi and Tomimatsu \cite{takahashi08} \cite[see also][]{pu20}. We do not solve the transverse force balance. Combining Eqs.~(\ref{eq:MHD_v}), (\ref{eq:MHD_eta}), (\ref{eq:MHD_E}), and (\ref{eq:MHD_L}) with $u_\mu u^\mu = -1$ leads to the Bernoulli equation,
\begin{equation}\label{eq:TTmethod}
  u_{\rm p}^2 = \frac{(\mathcal{E}-\mathcal{L}\Omega_{\rm F})^2 + f}{\beta^2 - f},
\end{equation}
where $u_{\rm p}^2 = u^r u_r + u^\theta u_\theta$,
\begin{equation}
  f \equiv g_{tt} + 2 g_{t\varphi}\Omega_{\rm F} + g_{\varphi\varphi}\Omega_{\rm F}^2 = -\alpha^2 + \gamma_{\varphi\varphi}(\Omega_{\rm F}-\Omega)^2,
  \label{eq:LSf}
\end{equation}
and
\begin{equation}
  \beta^2 \equiv \frac{H_{\varphi}^2}{B_{\rm p}^2},
\end{equation}
where $H_\varphi^2 = H^\varphi H_\varphi$ and $B_{\rm p}^2 = B^rB_r + B^\theta B_\theta$ (see Appendix~\ref{sec:app} for derivation of Eq.~\ref{eq:TTmethod}).
For given $\mathbf{B}_{\rm p}$, the appropriate set of $\Omega_{\rm F}, \mathcal{E},$ and $\mathcal{L}$ must be searched for obtaining the trans-magnetosonic flow solution as explained above, whereas the method of \cite{takahashi08} assumes that the function of $\beta^2$ is regular, for which one can obtain the trans-magnetosonic solution and $\mathbf{B}_{\rm p}$ for a given set of $\Omega_{\rm F}, \mathcal{E},$ and $\mathcal{L}$. In this method the regularity condition of the flow solution at the inner magnetosonic point is automatically satisfied just by the assumption on $\beta^2$. 
A convenient way is to set the form of $x \equiv D_{\rm p}^2/B_\varphi^2 = \gamma_{\varphi\varphi}(\Omega_{\rm F}-\Omega)^2/\beta^2$, where $D_{\rm p}^2 = D^r D_r + D^\theta D_\theta$ and $B_\varphi^2 = B^\varphi B_\varphi$, as
\begin{equation}
  x = \left(1+C\frac{\Delta}{\Sigma}\right)\left(\frac{\Omega - \Omega_{\rm F}}{\Omega_{\rm H} - \Omega_{\rm F}}\right)^2.
\end{equation}
This form of $x$ guarantees $\mathbf{D} = 0$ where $\Omega = \Omega_{\rm F}$ (see Eqs.~\ref{eq:consti} and \ref{eq:omegaF}) and the Znajek condition (Eq.~\ref{eq:znajek}) at the horizon, where $\Delta = 0$ and $\Omega = \Omega_{\rm H}$.
We take the numerical constant $C = -1$. Even for the other values of $C$, we find that the solution hardly changes, since $\Delta/\Sigma \ll 1$ in the inflow region. This form of $x$ was found to be consistent with that seen in a numerical solution that takes account of the transverse force balance \cite{ogihara21}.

\begin{figure}[t]
\begin{center}
  \includegraphics[scale=0.75]{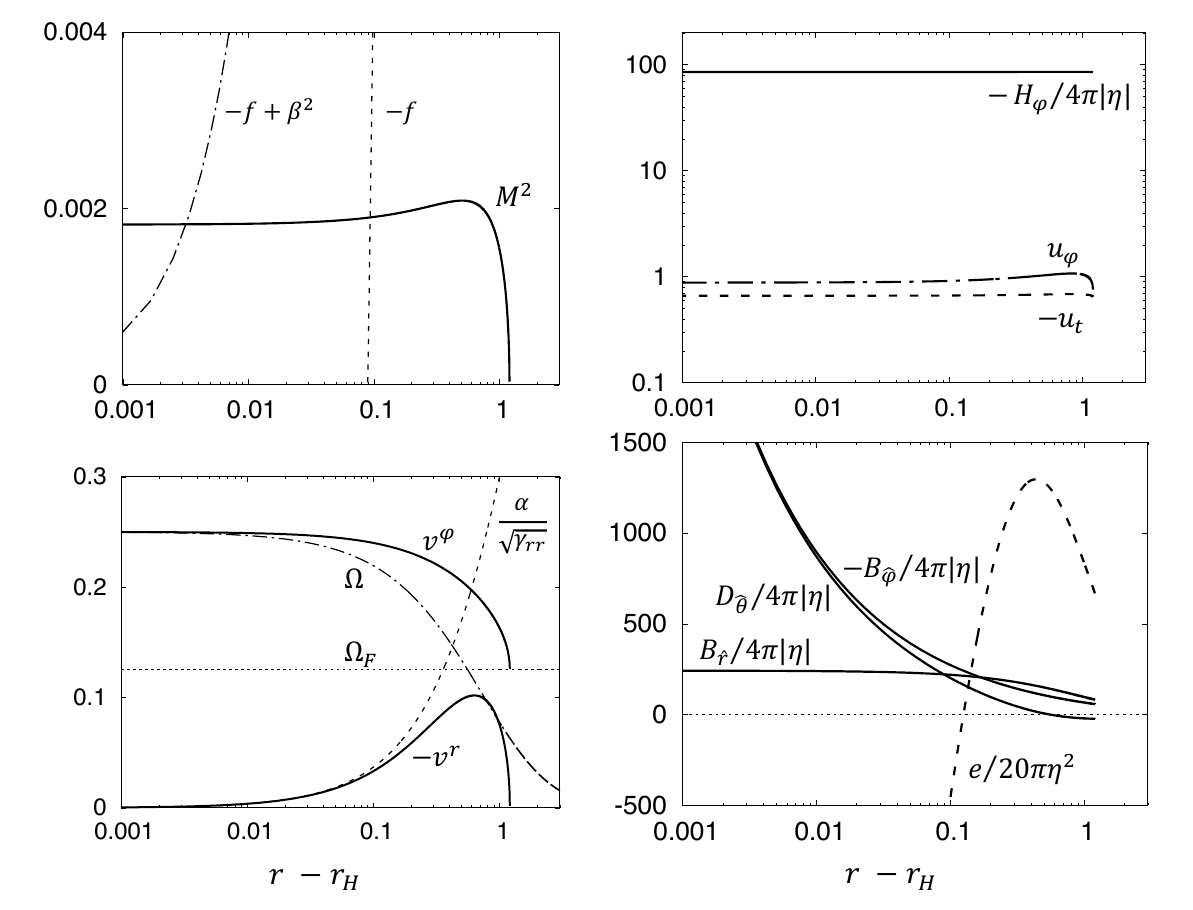}
\end{center}  
\caption{
  Solution of steady, axisymmetric MHD inflow towards the event horizon of a Kerr BH along a magnetic field line. The poloidal magnetic field is assumed to be in the split-monopole shape, i.e., $B^\theta = v^\theta = 0$. The parameters are $a=0.8$, $\theta = \pi/3$, $\Omega_{\rm F} = 0.5\Omega_{\rm H}$, $\mathcal{E} = -10$, and $\mathcal{L}\Omega_{\rm F} = -10.55$. The top left panel shows $M^2 = 4\pi \rho \alpha^2 u_{\rm p}^2/B_{\rm p}^2$ with $-f$ and $-f+\beta^2$ as functions of $r - r_{\rm H}$, while the top right panel $-H_\varphi/4\pi|\eta|$, $-u_t$, and $u_\varphi$.
The bottom left panel shows $-v^r$ and $v^\varphi$ with $\alpha/\sqrt{\gamma_{\varphi\varphi}}$, $\Omega$, and $\Omega_{\rm F}$ as functions of $r-r_{\rm H}$, while the bottom right panel $B_{\hat{r}}/4\pi|\eta|$, $-B_{\hat{\varphi}}/4\pi|\eta|$, $D_{\hat{\theta}}/4\pi|\eta|$, and $e/20\pi\eta^2$.
}
\label{fig:solution}
\end{figure}

We set the parameter values as consistent with the GRMHD simulation results introduced in Section~\ref{sec:intro}. The GRMHD simulations of RIAFs in the KS coordinates for $0<a<1$ with many different codes show similar quasi-steady structure of BH magnetospheres, which appears to be compatible with the current VLBI observational results \cite{nakamura18,porth19,narayan22,eht19b,eht21}. In those numerical solutions, $\Omega_{\rm F}/\Omega_{\rm H} \sim 0.3-0.5$, and the poloidal magnetic field looks like a split-monopole type particularly at $r \lesssim 5$ where the RIAF can be geometrically thin due to pressure of the magnetosphere \cite{tchekhovskoy10,penna13,ripperda22}. The parameter ranges of $\mathcal{E}$ and $\mathcal{L}$ will be explained later.

In Figure~\ref{fig:solution} we show an inflow solution for the parameters
$\Omega_{\rm F} = 0.5\Omega_{\rm H}$, $\mathcal{E} = -10$, and $\mathcal{L}\Omega_{\rm F} = -10.55$.
We have assumed the poloidal magnetic field as a split-monopole type, $B^\theta = 0$ (and thus $v^\theta = 0$ from Eq.~\ref{eq:MHD_v}), and have fixed $\theta = \pi/3$ for solving the Bernoulli equation.
The BH spin parameter is set to be $a=0.8$, for which the horizon radius and BH angular velocity are $r_{\rm H} \approx 1.57$ and $\Omega_{\rm H} \approx 0.255$, respectively. 
The upper left panel of Figure~\ref{fig:solution} shows Alfv\'{e}n Mach number $M^2 = 4\pi\rho\alpha^2 u_{\rm p}^2/B_{\rm p}^2$ as a function of $r - r_{\rm H}$. The inflow starts at the separation point, $r \simeq r_{\rm H} + 1.3$, passes the inner Alfv\'{e}n point where $M^2 = -f$ and the inner magnetosonic point where $M^2 = -f+\beta^2$, and continues towards the horizon.

The solution has $-u_t \ll -H_\varphi\Omega_{\rm F}/4\pi|\eta|$ and $u_\varphi \ll -H_\varphi/4\pi|\eta|$, as shown in the upper right panel of Figure~\ref{fig:solution}, which means that the entire inflow is magnetically dominated.
The matter energy $-u_t$ is nearly constant, that is, the matter appears to fall freely (see Section 2.1). $H_\varphi \approx {\rm const.}$ means that the poloidal electric current does not cross the magnetic field line (see Section~\ref{sec:ff}). The matter does not have negative energy, $-u_t > 0$, which is consistent with the GRMHD simulation results \cite{komissarov05}. Note that $u_t = u_\mu \xi^\mu$, $u_\varphi = u_\mu \chi^\mu$, $H_\varphi = {}^*F_{\mu\nu}\xi^\mu\chi^\nu$, and $\Omega_{\rm F} = -F_{t\theta}/F_{\varphi\theta}$ are the same in the BL and KS coordinates \cite{komissarov04a}.

The bottom left panel of Figure~\ref{fig:solution} shows the velocity field of the inflow.
At the separation point, $v^r = 0$ and $v^\varphi = \Omega_{\rm F}$. As approaching the horizon, 
\begin{equation}\label{eq:asymp}
  v^r \approx \frac{-\alpha}{\sqrt{\gamma_{rr}}}, ~~~ v^\varphi \approx \Omega_{\rm H}.
\end{equation}
These are the same asymptotic velocity components as a freely falling particle (Eq.~\ref{eq:v_ff}) \cite[see also][]{vanputten03}. The properties of electromagnetic fields are shown in the bottom right panel of Figure~\ref{fig:solution}. $B_{\hat{r}}$ is finite, while $D_{\hat{\theta}}$ and $B_{\hat{\varphi}}$ diverge. The sign of $D_\theta$ is the same as that of $E_\theta$ at the far zone, but changes at the point of $\Omega = \Omega_{\rm F}$. One also confirms $D^2 < B^2$ in the entire region. Note that the Znajek condition is approximately satisfied, $B_{\hat{\varphi}} \approx -D_{\hat{\theta}}$, at the inner magnetosonic point.\footnote{In the outflow, $B_{\hat{\varphi}} \approx E_{\hat{\theta}}$ at the outer magnetosonic point \cite[e.g.][]{lyubarsky09,toma13}.} The energy density of the electromagnetic field (Eq.~\ref{eq:EMenergy}) $e<0$ inside around the inner Alfv\'{e}n point
and diverges.
We note that the inner Alfv\'{e}n point is close to the inner light surface, $f=0$, which is generically inside the ergosphere \cite{komissarov04a}, and $e$ can be negative only inside the ergosphere \cite{komissarov09}.

Actually our solution of $B_{\hat{r}}$ deviates from the assumed monopole shape, but we confirm that the deviation is $< 12\%$ ($<18\%$ in the solution for $\theta = \pi/6$). Adjusting the assumed $\mathbf{B}_{\rm p}$ is required for obtaining exact solutions, but we consider that our solution is sufficient for understanding the detailed flow structure near the horizon and providing with the picture for the BH energy reduction mechanism (Section~\ref{sec:picture}).

\begin{figure}[t]
\begin{center}
  \includegraphics[scale=0.75]{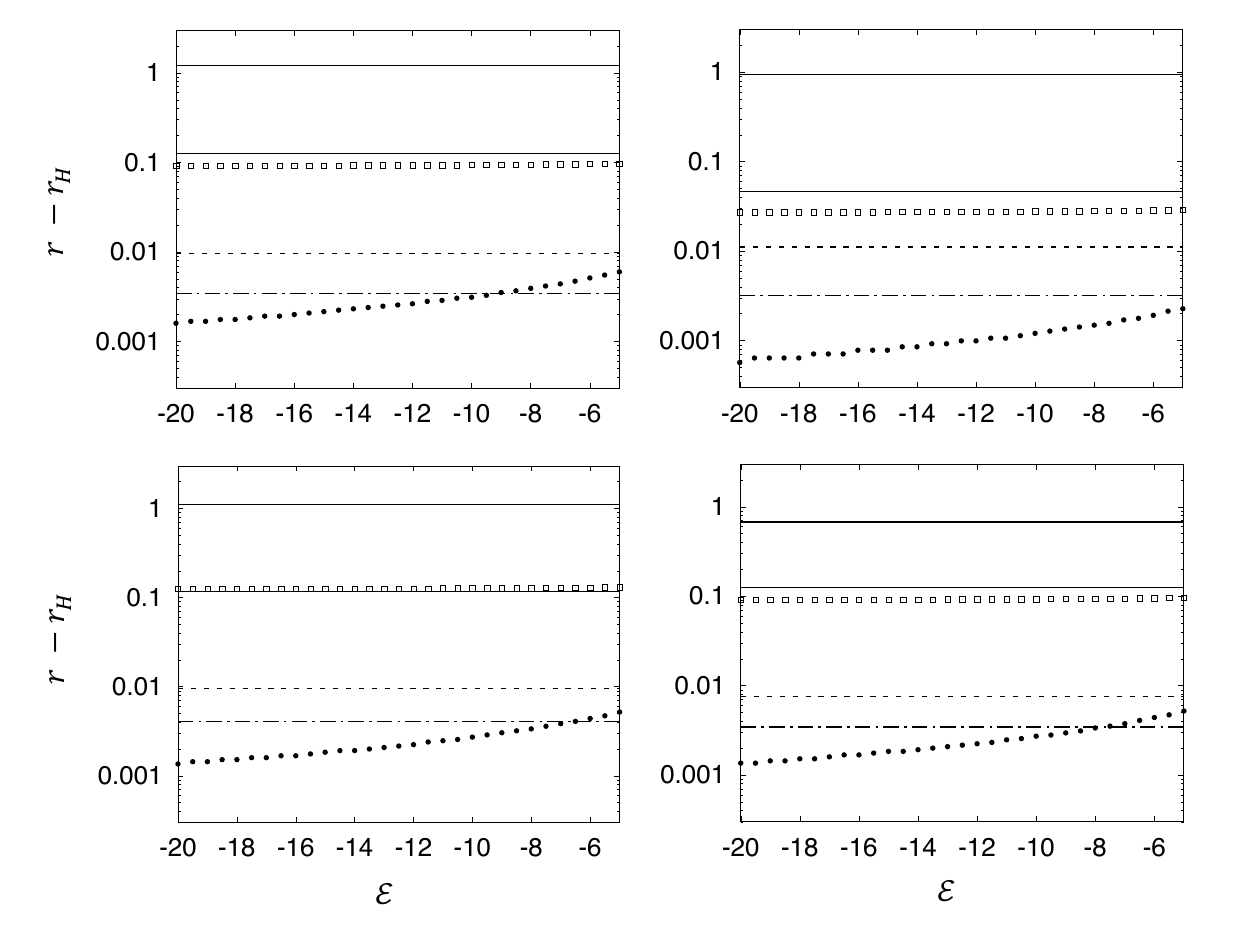}
\end{center}    
\caption{
  Characteristic radii of our MHD inflow solutions as a function of $\mathcal{E}$. For the top left panel, the parameter sets $(a, \Omega_{\rm F}/\Omega_{\rm H}, \mathcal{E}-\mathcal{L}\Omega_{\rm F})$ are $(0.8, 0.5, 0.55)$, which include the case of Figure~\ref{fig:solution}. We plot the radii of separation point (thick solid line), inner Alfv\'{e}n point (square), inner magnetosonic point (black dot), point of $e=0$ (thin solid line), that of $|v^r|/(\alpha/\sqrt{\gamma_{rr}}) = 0.99$ (dashed line), and that of $D_{\hat{\theta}}/|B_{\hat{\varphi}}| = 0.99$ (dot-dashed line) for the case of each $\mathcal{E}$. The parameter sets for the top right, bottom left, and bottm right panels are $(a, \Omega_{\rm F}/\Omega_{\rm H}, \mathcal{E}-\mathcal{L}\Omega_{\rm F}) = (0.5, 0.5, 0.55)$, $(0.8, 0.4, 0.55)$, and $(0.8, 0.5, 0.45)$, respectively.
}
\label{fig:radii}
\end{figure}

We also obtain solutions with other parameter value, $a= 0.5$, $\Omega_{\rm F}/\Omega_{\rm H} = 0.4$, or $\mathcal{E} - \mathcal{L}\Omega_{\rm F} = 0.45$ to confirm that they exhibit similar structures of inflows (Figure~\ref{fig:radii}). $\Omega_{\rm F}/\Omega_{\rm H} = 0.4$ is within the range, $\Omega_{\rm F}/\Omega_{\rm H} \sim 0.3-0.5$, which the GRMHD simulations indicate \cite{tchekhovskoy10,penna13}. For a given $\Omega_{\rm F}$, one has the maximum value of $-f$ from Eq.~(\ref{eq:LSf}), and then the existence of the separation point ($u_{\rm p}^2 = 0$ in Eq.~\ref{eq:TTmethod}) requires $\mathcal{E} - \mathcal{L}\Omega_{\rm F} < 0.58$ for $(a, \Omega_{\rm F}/\Omega_{\rm H}) = (0.8, 0.5)$. The cases of $(a, \Omega_{\rm F}/\Omega_{\rm H}) = (0.8, 0.4)$ and $(0.5, 0.5)$ have similar constraints, $\mathcal{E}-\mathcal{L}\Omega_{\rm F}<0.64$ and $< 0.72$, respectively. Figure~\ref{fig:radii} plots the radii of separation point (thick solid line), inner Alfv\'{e}n point (square), inner magnetosonic point (black dot), point of $e=0$ (thin solid line), that of $|v^r|/(\alpha/\sqrt{\gamma_{rr}}) = 0.99$ (dashed line), and that of $D_{\hat{\theta}}/|B_{\hat{\varphi}}|  = 0.99$ (dot-dashed line) for the case of each $\mathcal{E}$. $|\mathcal{E}| \gg 1$ is required for magnetically-dominated flows. As can be seen in Figure~\ref{fig:radii}, $e <0$ inside around the inner Alfv\'{e}n point, and $B_{\hat{\varphi}} \approx -D_{\hat{\theta}}$ around the inner magnetosonic point.
The ergosphere is inside the radius of  $r_{\rm es} - r_{\rm H} \approx 0.35$ for $a=0.8$ and $\approx 0.1$ for $a=0.5$, so that all the characteristic radii except the separation point are in the ergosphere.
We also confirm the same properties for the solutions with $\theta = \pi/6$ (see Appendix~\ref{sec:app2}).

\section{Mechanism of BH energy reduction}
\label{sec:picture}

As confirmed in Section~\ref{sec:mhd},
the matter energy $-u_t > 0$ in the inflow, so that the BZ process is not the same as the mechanical Penrose process.
Instead, the electromagnetic energy density (in the BL coordinates) $e < 0$ inside around the inner Alfv\'{e}n surface. Focusing on these properties, some theorists tried to find a picture how the BH reduces its rotational energy with infall of negative electromagnetic energy, but all of the proposed pictures were ambiguous \cite{komissarov09,takahashi90,koide14} (see Section~\ref{sec:literature}). Below we provide a clear picture for the BH energy reduction by the BZ process in the BL coordinates.

\begin{figure}[t]
  \begin{center}
    \includegraphics[scale=0.5]{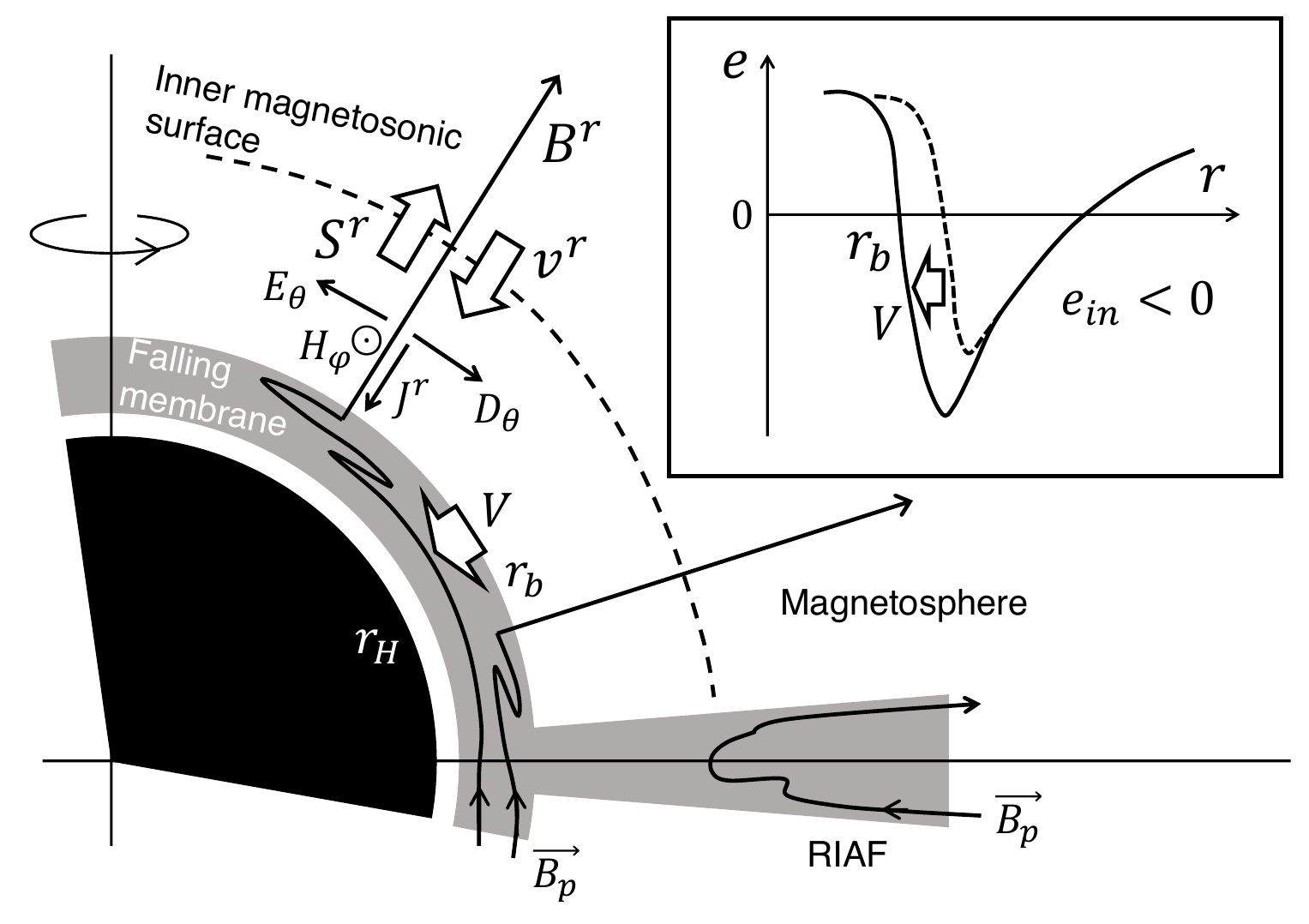}
  \end{center}
  \caption{
Schematic picture of the generic structure of the BH magnetosphere operating the BZ process, which magnifies the region close to the horizon for illustrative purpose. The magnetic pressure of the magnetosphere makes the RIAF thin near the horizon. The past accreted plasma or `falling membrane' with velocity $V < 0$ is followed by the magnetically-dominated inflow (like Fig.~\ref{fig:solution}) with $v^r < 0$. The magnetic field lines thread the falling membrane but not the horizon. The outward Poynting flux $S^r = E_\theta H_\varphi /4\pi\sqrt{\gamma}$ is produced at the boundary between the falling membrane and the inflow, which is associated with the inward extension of region of $e_{\rm in} < 0$ as illustrated in the top right panel. The poloidal electric current $J^r$ just flows inwards, maintaining $H_\varphi$, without crossing the magnetic field lines. 
  }
  \label{fig:schematic}
\end{figure}

BHs cannot have magnetic field by electric current inside themselves, unlike pulsars. The BH magnetosphere is formed by advection of the magnetic field frozen in the accretion disk (or RIAF) and disk wind, as shown in the GRMHD simulations \cite[e.g.][]{mckinney04,barkov08,tchekhovskoy11}. Therefore, the past accreted plasma threaded by the magnetic field should remain just above the horizon in the BL coordinates, as considered in the membrane paradigm \cite{thorne86} and illustrated in Figure~\ref{fig:schematic}. This past accreted plasma keeps falling towards the horizon, and we call it `falling membrane'. This holds waves and turbulence that have established the transverse force balance of ordered magnetic field in the magnetosphere. The magnetic field must not overtake the falling membrane under the MHD condition, so that it does not thread the event horizon, and no Poynting flux is directly released from the horizon. In the magnetosphere with the established ordered field, the steady structure is formed between the inner and outer magnetosonic surfaces, as explained in Section 3, and the inflow $(v^r < 0)$ which carries the electromagnetic field $E_\theta$ and $H_\varphi$ (forming the outward Poynting flux $S^r = E_\theta H_\varphi/4\pi\sqrt{\gamma}$)
is following the falling membrane.
The inflow inside the inner magnetosonic surface keeps extending towards the event horizon, and in this sense, the front of the inflow is not steady.

To clarify the mechanism of BH energy reduction, we consider the Amp\`{e}re-Maxwell equation (in Eq.~\ref{eq:maxwell2}) at the boundary between the falling membrane and the inflow. Let us assume that the poloidal magnetic field of the inflow is of split-monopole shape. The boundary is just above the horizon, so that it is spherical. We define its radius by $r_{\rm b}$ and its coordinate radial velocity by $dr_{\rm b}/dt = V$. Then $D^\theta$ and $H_\varphi$ can be written by \cite[cf.][]{toma16}
\begin{equation}
  D^\theta = D^\theta_{\rm mb} H(-R) + D^\theta_{\rm in} H(R), 
\end{equation}
\begin{equation}
  H_\varphi = H_\varphi^{\rm mb} H(-R) + H_\varphi^{\rm in} H(R),
\end{equation}
where $H(R)$ is a Heaviside step function with
$R \equiv r - r_{\rm b}$.
Here and hereafter the quantities with subscript or superscript `mb' represent the field components in the falling membrane, while those with `in' represent the field components given by the calculation result of the steady, axisymmetric inflow as shown in Figure~\ref{fig:solution}.
Substituting these equation into the Amp\`{e}re-Maxwell equation (in Eq.~\ref{eq:maxwell2}) leads to
\begin{eqnarray}
  &&-D^\theta_{\rm mb} V \delta(R) + D^\theta_{\rm in} V \delta(R)
  -\frac{1}{\sqrt{\gamma}}(\partial_r H_\varphi^{\rm mb}) H(-R)
  +\frac{1}{\sqrt{\gamma}}H_\varphi^{\rm mb}\delta(R) \nonumber\\
  &&+\frac{1}{\sqrt{\gamma}}(\partial_r H_\varphi^{\rm in}) H(R)
  -\frac{1}{\sqrt{\gamma}}H_\varphi^{\rm in}\delta(R)
  = 4\pi \eta^\theta\delta(R),
\end{eqnarray}
where $\eta^\theta$ is possible surface electric current.
This gives us for $R \to 0$
\begin{equation}\label{eq:stepMaxwell}
  H_\varphi^{\rm in} - H_\varphi^{\rm mb} = \sqrt{\gamma}V(D^\theta_{\rm in} - D^\theta_{\rm mb}) - 4\pi \sqrt{\gamma}\eta^\theta.
\end{equation}
The membrane is falling by the dominant gravitational force, so that one has (Eq.~\ref{eq:v_ff}),
\begin{equation}
  V \approx \frac{-\alpha}{\sqrt{\gamma_{rr}}}.
\label{eq:V}
\end{equation}
This is the same as the asymptotic radial velocity of the inflow (Eq.~\ref{eq:asymp}).

Generally the relation $B_{\hat{\varphi}} \approx -D_{\hat{\theta}}$ is satisfied at $r \approx r_{\rm H}$ (Eq.~\ref{eq:znajek}), which can be rewritten as
\begin{equation}
  \frac{H_\varphi}{\alpha\sqrt{\gamma_{\varphi\varphi}}} \approx -\sqrt{\gamma_{\theta\theta}}D^\theta.
  \label{eq:znajek_Hphi}
\end{equation}
Thus this relation is valid for the electromagnetic fields in the falling membrane. We have seen that this relation is also valid inside around the inner magnetosonic surface in the inflow (Section~\ref{sec:mhd}).
Therefore, by using Eq.~(\ref{eq:V}) valid for both sides of the boundary, we obtain 
\begin{equation}\label{eq:relation}
  H_\varphi \approx \sqrt{\gamma} V D^\theta,
\end{equation}
{\it at both sides of the boundary.} This implies in Eq.~(\ref{eq:stepMaxwell})
\begin{equation}
  \eta^\theta \approx 0,
\end{equation}
that is, the electric current does not flow across the magnetic field lines. The Znajek condition means that $H_\varphi^{\rm in}$ is not produced by the conductive current on the membrane but is associated with the displacement current $\sqrt{\gamma} V D^\theta_{\rm in}$. This also means that the poloidal electric current which maintains $H_\varphi^{\rm in}$ does not form a closed circuit inside the BH magnetosphere.

The electromagnetic energy density of the inflow (Eq.~\ref{eq:EMenergy}) at $r \approx r_{\rm H}$ is written by taking account of our assumption $B^\theta = 0$ and Eq.~(\ref{eq:consti}) \cite[cf.][]{komissarov09} as well as the condition inside around the inner magnetosonic surface, $B^r_{\rm in} B_r^{\rm in} \ll B^\varphi_{\rm in}B_\varphi^{\rm in} \approx D^\theta_{\rm in}D_\theta^{\rm in}$, 
\begin{eqnarray}
  e_{\rm in} &&= \frac{1}{8\pi\alpha}\left[\alpha^2 (B^r_{\rm in}B_r^{\rm in} + B^\varphi_{\rm in} B_\varphi^{\rm in}) + \gamma_{\varphi\varphi}(\Omega_{\rm F}^2 - \Omega^2)B^r_{\rm in}B_r^{\rm in}\right] \nonumber\\
  &&\approx \frac{\gamma_{rr}\gamma_{\varphi\varphi}}{4\pi \alpha}\Omega_{\rm F}(\Omega_{\rm F}-\Omega_{\rm H})(B^r_{\rm in})^2
  = \frac{-\sqrt{\gamma}}{4\pi}\Omega_{\rm F}D^\theta_{\rm in}B_{\rm in}^r < 0, 
\end{eqnarray}
so that one can derive by using Eq.~(\ref{eq:relation})
\begin{equation}\label{eq:SreV}
  S_{\rm in}^r = \frac{-\Omega_{\rm F}H_\varphi^{\rm in}}{4\pi}B^r_{\rm in} = e_{\rm in} V.
\end{equation}
Therefore, the Poynting flux is produced at the boundary between the falling membrane and the inflow, not at the event horizon, and its production is associated with the inward extension of region with $e_{\rm in} < 0$, as illustrated in the top right panel of Figure~\ref{fig:schematic}. The creation of $e_{\rm in} < 0$ is equivalent to reduction of the rotational energy of the Kerr spacetime. We note that $e_{\rm in}$ diverges but $V$ converges to zero. The sign of the electromagnetic energy density of the falling membrane can be either positive or negative, which is not relevant for the mechanism of BH energy reduction.

In our picture, the ergosphere plays the role of making $e_{\rm in} < 0$.
Let us recall that the BZ process is the steady production of the outward Poynting flux along the magnetic field lines threading the horizon. This steady process is established between the inner and outer magnetosonic surfaces. Thus the front of the inflow is already inside the inner magnetosonic surface, where $e_{\rm in} < 0$, since the inner magnetosonic surface is well inside the ergosphere. Consequently, the spacetime reduces its rotational energy by the inward movement of the inflow front.

\section{Discussion}
\label{sec:discussion}

In this paper, we have treated the membrane of past accreted matter that should exist above the event horizon and falls with velocity $V \approx -\alpha/\sqrt{\gamma_{rr}}$ in the BL coordinates. We have clarified the generic existence of unsteady front of the magnetically-dominated inflow where the outward Poynting flux $S_{\rm in}^r = E_\theta^{\rm in} H_\varphi^{\rm in}/4\pi\sqrt{\gamma}$ and negative electromagnetic energy $e_{\rm in}$ are created (see Eq.~\ref{eq:SreV}). The creation of $e_{\rm in} < 0$ is equivalent to the rotational energy reduction of the Kerr spacetime. Most of the previous studies except the membrane paradigm \cite{thorne86} assume that the magnetic field lines thread the event horizon even in the BL coordinates \cite[e.g.][]{blandford77,takahashi90,takahashi08,beskin00,huang20}. The situation without the falling membrane is simple for searching the MHD flow solutions, but unrealistic and not suitable for understanding the mechanism of BH energy reduction.

We have also shown that the Znajek condition reflects the association of $H_\varphi^{\rm in}$ and the displacement current $\sqrt{\gamma} V D^\theta_{\rm in}$ (see Eq.~\ref{eq:relation}), and that the poloidal electric current maintaining $H_\varphi$ does not form a closed circuit within the magnetosphere.

In the following, we discuss the previous studies on the mechanism of BH energy reduction and the current driving force in the BZ process.

\subsection{Previous studies on the BH energy reduction}
\label{sec:previous}

\subsubsection{The membrane paradigm}

The membrane paradigm considers that the magnetic field lines are threading the membrane of past accreted matter similarly to that illustrated in Figure~\ref{fig:schematic} \cite{thorne86}. Since the matter falls with velocity $V \approx -\alpha/\sqrt{\gamma_{rr}} \to 0$ for $r \to r_{\rm H}$, the membrane paradigm assumes that the membrane is at a fixed radius, i.e., $V = 0$, and that the origin of $H_\varphi^{\rm in}$ is conductive surface electric current,
\begin{equation}
  \sqrt{\gamma_{rr}}\eta^\theta = \frac{1}{4\pi}\alpha D^\theta.
  \label{eq:ohm}
\end{equation}
Indeed, Eq.~(\ref{eq:ohm}) is derived from Eqs.~(\ref{eq:stepMaxwell}) and (\ref{eq:znajek_Hphi}) by setting $V=0$ and $H_\varphi^{\rm mb} = 0$. In the paradigm, Eq.~(\ref{eq:ohm}) is interpreted as Ohm's law, and the conductive surface current is considered to close the poloidal electric current circuit and to reduce the BH energy by $\mathbf{J} \times \mathbf{B}$ force.

In reality, however, the membrane keeps falling with convergently small velocity $V$, and $\eta^\theta \approx 0$ as shown in Section 4. $H^{\rm in}_\varphi$ is not formed by the cross-field current $\eta^\theta$, but determined by the conditions at the inner and outer magnetosonic surfaces (or the frame-dragging effect between the surfaces), as explained in Section~\ref{sec:mhd}.
Inside the membrane the electromagnetic field is arbitrary, depending on the past accretion history \cite[see also][]{punsly89}.
The membrane paradigm is misleading with respect to the origin of the Poynting flux, the BH energy reduction mechanism, and the structure of poloidal electric current.

\subsubsection{Different views in the BL and KS coordinates}

Toma and Takahara \cite{toma16} discussed the time-dependent behavior of the front of magnetically-dominated inflow, but they assumed a vacuum between the horizon and the inflow, instead of the falling membrane. The existence of the falling membrane is more natural than the vacuum. Although they provided with the key equations, Eqs.~(\ref{eq:maxwell2}) and (\ref{eq:e_cons}), at the front of the inflow, they did not find the mechanism of BH energy reduction by formation of $e_{\rm in} < 0$. This was partly because they insisted on seeking properties invariant between the BL and KS coordinates, and $e_{\rm in}$ is positive in the KS coordinates. 

\subsubsection{Infall of negative energy objects}
\label{sec:literature}

The problem what kind of negative energy falls into the BH in the BZ process was first discussed in depth by Komissarov \cite{komissarov09} \cite[see also][]{lasota14}. The argument there utilized the BL coordinates and tried to find a certain equation $S^r = e v^r_{\rm EM}$
for the part of the inflow with $D^\theta > 0$ (i.e., $\Omega > \Omega_{\rm F}$)
instead of only the front of the inflow, but the infall velocity $v^r_{\rm EM}$ was not identified.
It was emphasized that the electromagnetic energy flux measured by the FIDOs, $S^r_{\rm FIDO} = D_\theta B_\varphi/4\pi\sqrt{\gamma}$, is in the opposite direction of $S^r = E_\theta H_\varphi/4\pi\sqrt{\gamma}$ where $\Omega > \Omega_{\rm F}$. However, $S^r_{\rm FIDO}$ is not straighforwardly related to the electromagnetic energy density in the coordinate basis $e$ (`energy density measured at infinity'). In contrast, Eq.~(\ref{eq:SreV}) is constituted of only quantities in the coordinate basis, $S^r_{\rm in}, e_{\rm in},$ and $V$.

Koide and Baba \cite{koide14} wrote $S^{\hat{r}} = \alpha e v^r_{\rm EM}$ at the horizon similarly to Eq.~(\ref{eq:SreV}), but they interpreted $v^r_{\rm EM}$ as the steady particle drift velocity measured by the FIDOs, $D_{\hat{\theta}} B_{\hat{\varphi}}/B^2$. This interpretation is similar to that in \cite{komissarov09}.

Takahashi et al. \cite{takahashi90} argued that $S^r \approx \mathcal{E} \alpha \rho u^t v^r$ (Eq.~\ref{eq:MHD_E}) means a negative energy infall with $\mathcal{E} < 0$ and $v^r < 0$ for the entire inflow. However, $\mathcal{E}$ is related to $e$ only at the front of inflow.
Moreover, all the above interpretations in \cite{komissarov09,koide14,takahashi90} tried to find an equation of type $S^r = e v^r_{\rm EM}$ for the steady part of the inflow. However, $S^r$ is the product of the steady fields $E_\theta$ and $H_\varphi$, so that its velocity could not be defined. In our picture, we have considered the velocity $V$ of the inflow front, which is unsteady part of the inflow.


Kinoshita and Igata \cite{kinoshita18} developed the formalism of world sheets of rigidly rotating Nambu-Goto strings, which has properties similar to the field sheets of force-free electromagnetic fields, and suggested that the `stretching of the magnetic field lines' plays the role of the infalling objects. This might correspond to the front of the MHD inflow in our picture.

An analogy of the Alfv\'{e}n wave superradiance to the BZ process has also been investigated \cite{noda22}. It is beyond the scope of this paper to proceed with the discussion in this direction.

\subsection{Driving force of poloidal current}

The remaining problem is what drives the poloidal electric current maintaining $H_\varphi$.
In the unipolar induction of the pulsar winds, the $\mathbf{v} \times \mathbf{B}$ force in the matter-dominated star drives the poloidal current in the direction of $-\mathbf{E}$, which drives the steady poloidal current circuit. In contrast, the BZ process has no matter-dominated region, and the cross-field current cannot flow effectively in the inflow or the outflow as seen in Section~\ref{sec:mhd}. The charged particles carrying the current are provided from the outside into the ordered magnetic field of the BH magnetosphere \cite[e.g.][]{levinson11,kimura22}, which must be adjusted to carry the poloidal current and maintain $H_\varphi$. 

Suppose that the conductive cross-field current (Eq.~\ref{eq:ohm}) was flowing on the membrane as in the membrane paradigm, then one would consider that the cross-field current in the opposite direction is driven somewhere, for example at the separation surface or the surface of $\Omega = \Omega_{\rm F}$ \cite{okamoto05}. However, we have shown that $\eta^\theta \approx 0$ on the falling membrane, and thus no cross-field current is required to be driven. The magnetically dominated inflow and outflow has no circuit of poloidal electric current, and the poloidal current just flows along the magnetic field lines.

This poloidal current could keep adjusted by weak violations of ideal MHD conditions \cite{komissarov09}. Indeed, the recently developed general relativistic particle-in-cell simulations show that $\mathbf{D} \cdot \mathbf{B} = 0$ is weakly violated everywhere on the plasma skin depth scale \cite{levinson18,parfrey19} and strong violation can occur quasi-periodically at the point of $\Omega = \Omega_{\rm F}$ as shown in the 1-dimensional simulations \cite{chen20,kisaka20,kin24} or at the inner light surface as shown in the 2-dimensional simulations \cite{crinquand20,crinquand21}. It is important to investigate the real driving site, and its observational signature would provide an evidence of the BZ process \cite[e.g.][]{levinson11}.

Since the poloidal electric current is not closed within the BH magnetosphere, the BH has a possibility of charging up. Komissarov \cite{komissarov22} shows that the BZ process can operate even by charged BHs. The matter-dominated accretion disk around the equatorial plane will drive the return current, which prevents the BH from charging up. For the magnetically-arrested disk phase, the BH could be charged up transiently. Magnetic reconnection of the magnetospheric toroidal field around the equatorial plane \cite{ripperda22,kimura22,crinquand21} will contribute to the return current.


\section*{Acknowledgment}

K.~T. reports with sadness that our coauthor Fumio Takahara, who contributed seminal ideas to this work, passed away 2023 June 6. K.~T. thanks T.~Harada, M.~Kimura, A.~Naruko, K.~Ogasawara, S. J.~Tanaka, S.~Kisaka, and K.~Kin for useful discussions. This work is partly supported by JSPS Grants-in-Aid for Scientific Research No.~18H01245 (K.~T.) and No.~24K07100 (M.~N.).

\appendix

\section{Derivation of Eq.~(\ref{eq:TTmethod})}
\label{sec:app}

Combining Eqs.~(\ref{eq:MHD_v}), (\ref{eq:MHD_eta}), (\ref{eq:MHD_E}), and (\ref{eq:MHD_L}), $u_t, u_\varphi,$ and $H_\varphi/4\pi\eta$ can be described by $\Omega_{\rm F}, \mathcal{E}, \mathcal{L},$ and $M^2$,
\begin{equation}\label{eq:app_ut}
  u_t = \frac{G_t(\mathcal{L}\Omega_{\rm F} - \mathcal{E})-M^2\mathcal{E}}{M^2 + f},~~~
  u_\varphi = \frac{G_\varphi(\mathcal{L}\Omega_{\rm F} - \mathcal{E})+M^2\mathcal{L}}{M^2 + f},
\end{equation}
\begin{equation}\label{eq:app_Hphi}
  \frac{H_\varphi}{4\pi\eta} = \frac{-G_\varphi\mathcal{E} - G_t\mathcal{L}}{M^2 + f},
\end{equation}
where $G_t \equiv g_{tt} + g_{t\varphi}\Omega_{\rm F}$ and $G_\varphi \equiv g_{t\varphi} + g_{\varphi\varphi}\Omega_{\rm F}$ are defined.
Substituting Eq.~(\ref{eq:app_ut}) into $u_\mu u^\mu = -1$ leads to
\begin{equation}
  \frac{1}{(M^2+f)^2}\left[\epsilon^2 (f+2M^2) - \frac{k}{\alpha^2\gamma_{\varphi\varphi}}M^4\right] + u_{\rm p}^2 = -1,
\end{equation}
where $\epsilon \equiv \mathcal{E} - \mathcal{L}\Omega_{\rm F}$ and $k \equiv g_{\varphi\varphi} \mathcal{E}^2 + 2g_{t\varphi}\mathcal{E}\mathcal{L} + g_{tt}\mathcal{L}^2$.
By multiplying both sides of this equation by $f$, and using $\alpha^2\gamma_{\varphi\varphi} \epsilon^2+fk = (G_\varphi \mathcal{E} + G_t \mathcal{L})^2$ and Eq.~(\ref{eq:app_Hphi}), one obtains
\begin{equation}
  \epsilon^2 - \frac{H_\varphi^2}{(4\pi\eta\alpha)^2}M^4 + f u_{\rm p}^2 = -f.
\end{equation}
This is rewritten as Eq.~(\ref{eq:TTmethod}).

These equations include $\eta$ only as $H_\varphi/\eta$ and $B_{\rm p}/\eta$ since $M^2 = 4\pi \rho \alpha^2 u_{\rm p}^2/B_{\rm p}^2 = \alpha u_{\rm p}/(B_{\rm p}/4\pi\eta)$. Therefore $\eta$ does not affect the structure of flow solution, but just affects the normalizations of field strengths.

\section{MHD inflow Solutions for $\theta = \pi/6$}
\label{sec:app2}

\begin{figure}[t]
\begin{center}
  \includegraphics[scale=0.75]{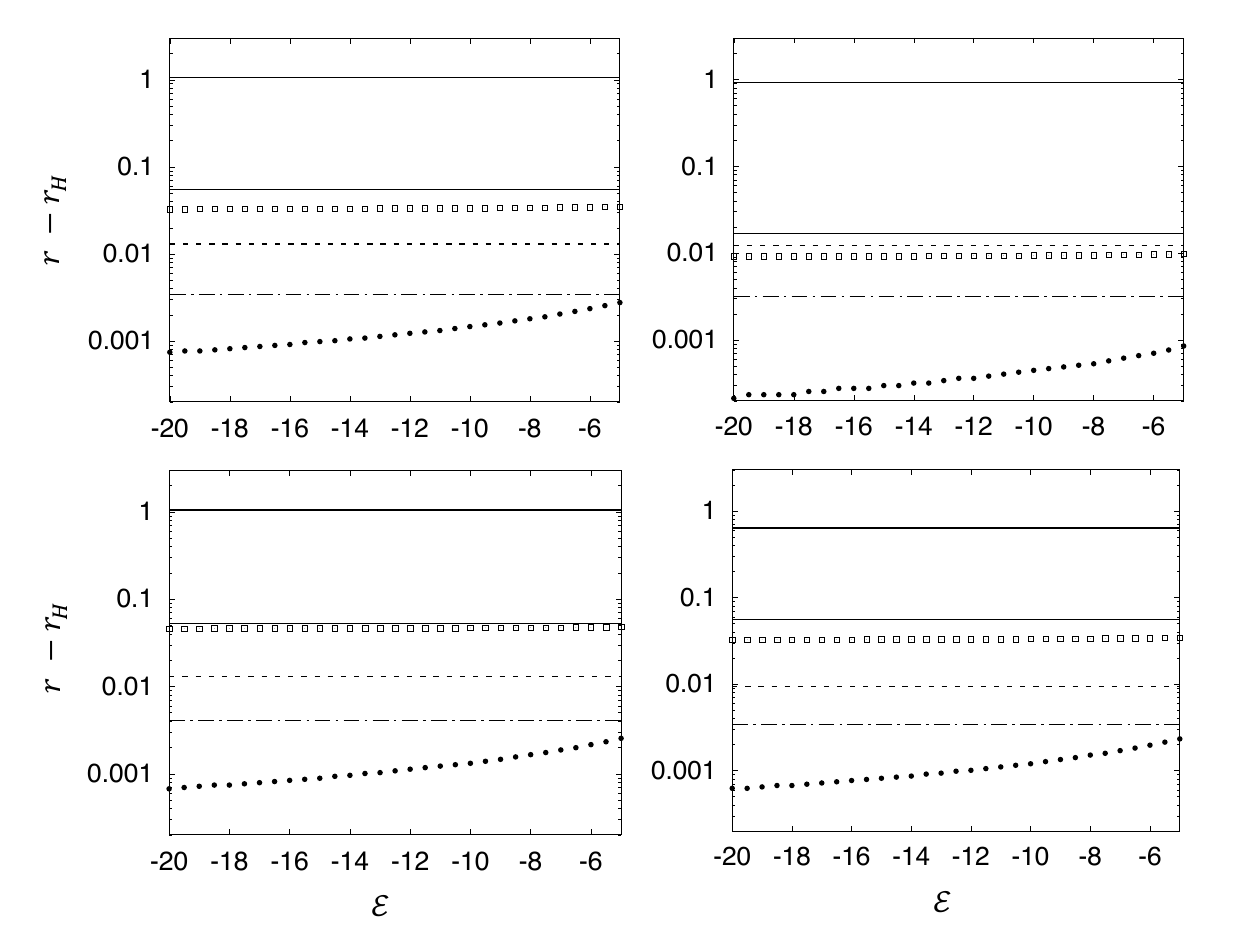}
\end{center}    
\caption{
  Characteristic radii of our MHD inflow solutions for $\theta = \pi/6$. We plot the radii of separation point (thick solid line), inner Alfv\'{e}n point (square), inner magnetosonic point (black dot), point of $e=0$ (thin solid line), that of $|v^r|/(\alpha/\sqrt{\gamma_{rr}}) = 0.99$ (dashed line), and that of $D_{\hat{\theta}}/|B_{\hat{\varphi}}| = 0.99$ (dot-dashed line) for the case of each $\mathcal{E}$. The parameter sets for the top left, top right, bottom left, and bottm right panels are $(a, \Omega_{\rm F}/\Omega_{\rm H}, \mathcal{E}-\mathcal{L}\Omega_{\rm F}) = (0.8, 0.5, 0.55)$, $(0.5, 0.5, 0.55)$, $(0.8, 0.4, 0.55)$, and $(0.8, 0.5, 0.45)$, respectively.
}
\label{fig:radii_app}
\end{figure}

For reference, we show the characteristic radii of the steady, axisymmetric, cold MHD inflow solutions along a split-monopole poloidal magnetic field line with $\theta = \pi/6$ in Figure~\ref{fig:radii_app}. For $\theta = \pi/6$, the ergosphere is inside the radius of $r_{\rm es} - r_{\rm H} \approx 0.15$ for $a=0.8$ and $r_{\rm es} \approx 0.03$ for $a=0.5$.

\let\doi\relax

\end{document}